\documentclass[aps,prb,twocolumn]{revtex4}
\usepackage{amsmath}
\usepackage{amssymb}
\usepackage{graphicx}%
\usepackage{amsfonts}

\begin{document}
\title{Full counting statistics of super-Poissonian shot noise in multi-level
  quantum dots}
\author{W. Belzig} 
\affiliation{
  Departement Physik und Astronomie, Universit\"at Basel,
  Klingelbergstrasse~82,
  CH-4056 Basel, Switzerland}


%% \pacs{72.70.+m}{Noise processes and phenomena}
%% \pacs{73.23.-b}{Electronic transport in mesoscopic systems}
%% 73.50.Td Noise processes and phenomena

\begin{abstract}
We examine the full counting statistics of quantum dots, which display
super-Poissonian shot noise. By an extension to a generic
situation with many excited states we identify the underlying transport
process. The statistics is a sum of independent Poissonian processes of
bunches of different sizes, which leads to the enhanced noise. The obtained
results could be useful to determine transport characteristics in
molecules and large quantum dots, since the noise (and higher cumulants)
allow to identify the internal level structure, which is not visible
in the average current.
\end{abstract}

\maketitle

Current fluctuations in mesoscopic structure far from equilibrium
provide a great deal of insight into the relevant transport mechanism
(see Refs.  \cite{blanter:00} and \cite{nazarov:03} for recent reviews).
A complete understanding, however, requires to go beyond the
second-order correlation function ('shot noise') and to study the full
counting statistics \cite{levitov:93}, which yields all zero-frequency
current-correlation functions at once. In contrast to transport
of classical particles, which leads to Poisson statistics of the
transfered charge, quantum transport of electrons is described by a
binomial statistics \cite{levitov:93}, which explains the suppression of
noise in non-interacting conductors \cite{khlus:87}. One way to observe
super-Poissonian noise is to use a superconducting injector, which
yields an enhanced noise \cite{2e} as result of the doubled charge
transfer due to Andreev reflection \cite{2efcs}. In systems with two
superconducting leads even larger charge transfers can occur
\cite{cuevas:03}. On the other hand, repulsive interactions as for
example encountered in simple quantum dots can lead to a suppressed shot
noise \cite{setnoise,bagrets:03,thielmann:03}, similar to noninteracting double
tunnel junction systems.

Some recent works found, surprisingly, a strong enhancement of the shot
noise in the transport through interacting quantum dots, in which the
spin degeneracy was lifted \cite{bulka,cottet}. A similar enhancement was
found in systems of multiple coupled quantum dots \cite{coupled}, in 
resonant tunneling diodes \cite{blanter:99} or due to cotunneling
\cite{sukhorukov}. In addition, in multi-terminal quantum dots the
enhanced noise results in positive cross correlations \cite{cottet}, which is
probably the simplest mechanism to produce unusual positive cross
correlations in a fermionic circuit \cite{buettiker:92}. We note here
that these mechanisms do not require a negative differential resistance,
which has been discussed in a number of works as possible source of
enhanced shot noise \cite{negdiff}.

In this article we examine the full counting statistics of transport
through a quantum dot, which contains several levels. We limit the
discussion to the sequential tunneling regime and for the moment to two
energy levels $\epsilon_{\pm}$, which in equilibrium are below the Fermi
energy of the leads. Under the condition of strong Coulomb blockade,
viz. the charging energy $E_c \gg \epsilon_+-\epsilon-$, the occupation
of a single level may block the transport through the other level.  This
mechanism was termed \textit{dynamical channel blockade} and can lead to
a strongly enhanced noise \cite{cottet,cottet:prb} (related
mechanisms, due to coupling to phonons \cite{oppen} or bistable systems
\cite{jordan}, were recently discussed). Below we will show that the
strongly enhanced Fano factor is a result of a combination of many
different Poissonian processes, which transfer multiple charges. It
follows from an analysis of the full counting statistics that these
transport properties can only be understood from a study of all
cumulants of the current. These transport properties are markedly
different from those in the situation, in which the levels are situated
above the Fermi energy in equilibrium.  Here, the fluctuations are
similar to the noninteracting case and the noise is always
sub-Poissonian.

To study the full counting statistics, we recall here that we are
dealing with the probability $P(N)$ that $N$ charges are transfered in a
period $t_0$, which is related to the cumulant generating function
$S(\chi)$ as
\begin{equation}
  \label{eq:cgf}
  e^{-S(\chi)}=\sum_N P(N) e^{iN\chi}\,.
\end{equation}
The knowledge of the CGF is equivalent to the knowledge of the
counting statistics. From the CGF we can obtain cumulants
$C_n=-(-i\partial_\chi)^n S(\chi)|_{\chi=0}$, which we identify with the
current $\bar I=eC_1/t_0$, the noise $S_I=2e^2 C_2/t_0$ and similar for
higher cumulants.

\begin{figure}[tbp]
  \centering
  \includegraphics[width=8cm,keepaspectratio,clip]{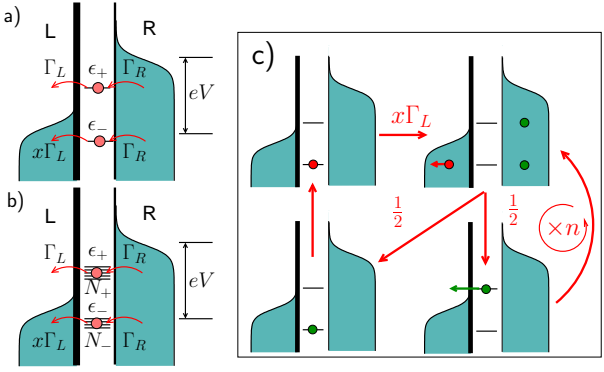}
  \caption{a) Two-terminal quantum dot with two levels. The tunneling
    rate from the lower level to the left lead is thermally activated
    due to finite occupation of the left lead. All other
    tunneling processes have much larger rates. Coulomb charging allows
    only one of the levels to be occupied. b) The same with $N_+$ upper
    levels and $N_-$ lower levels. c) Transport cycles leading to
    super-Poissonian shot noise. With a small tunneling rate $x\Gamma_L$
    a cycle is initiated by an electron tunneling from the dot to the
    left lead, followed by $n$ cycles ($n=0,1,2,\ldots$) of tunneling through the
    upper level with probability $1/2$. Finally the process is stopped
    by a tunneling event from the right terminal to the lower level with
    probability $1/2$. In total $n+1$ charges are transfered in a cycle
    with a rate $x\Gamma_L (1/2)^{n+1}$, which explains the
    super-Poissonian noise and all other transport characteristics.}
  \label{fig:cycles}
\end{figure}

The simplest system, which exhibits the phenomenon, is a two-terminal
quantum dot with more than one level in the situation shown in
Fig.~\ref{fig:cycles}a). In fact, this is the situation, which was first
studied in \cite{cottet:prb}. The applied bias voltage and the level
configuration is such that the level $\epsilon_-$ is located below
chemical potentials of both leads and the level $\epsilon_+$ in
between. The charging energy is so large that only one of the levels can
be occupied at the same time.

In this limit the system is described by a Master equation of the form
\begin{equation}
  \frac{d}{dt}\left(
    \begin{array}[c]{c}
      P_-\\P_+\\P_0
    \end{array}\right)
  =
  \left(
  \begin{array}[c]{ccc}
    -x\Gamma_L & 0 & \Gamma_R\\
    0 & -\Gamma_L & \Gamma_R\\
    x\Gamma_L & \Gamma_L & -2 \Gamma_R
  \end{array}\right)
  \left(
  \begin{array}[c]{c}
      P_-\\P_+\\P_0
    \end{array}\right)
  \equiv \hat M \vec P.
\end{equation}
Here $P_\pm$ describes the occupation of the upper/lower level and $P_0$
is the probability that the island is empty. The rates are parameterized
by the bare tunneling rates $\Gamma_{L/R}$ to tunnel out of the dot to
the left or to tunnel onto the dot from the right. The (small) parameter
$x$ is determined by the thermal occupations in the left lead: tunneling
from the lower state is suppressed by a factor $x=1-f(\epsilon_-)$,
since the states in the left lead are mostly occupied.

Now we determine the counting statistics using the method of Bagrets
and Nazarov \cite{bagrets:03}. We add the corresponding counting
factors $\exp(i\chi)$ to the entries of the Master equation, which
involve changes of the dot occupation and refer to tunneling into
one of the terminals. This procedure leads to the rate matrix
\begin{equation}
  \hat M(\chi)=  \left(
  \begin{array}[c]{ccc}
    -x\Gamma_L & 0 & \Gamma_R\\
    0 & -\Gamma_L & \Gamma_R\\
    x\Gamma_Le^{i\chi} & \Gamma_Le^{i\chi} & -2 \Gamma_R
  \end{array}\right)\,,
\end{equation}
where we have chosen to count the charges in the left terminal. The
counting statistics is determined from the lowest eigenvalue
$\lambda_0(\chi)$ of $\hat M(\chi)$ according to
\begin{equation}
  S(\chi)=-\lambda_0(\chi) t_0\,.
\end{equation}
The eigenvalue equation reads
\begin{eqnarray}
   \nonumber
  0 & = & (x\Gamma_L+\lambda) 
  \left[\Gamma_L\Gamma_R
    e^{i\chi}-(\Gamma_L+\lambda)(2\Gamma_R+\lambda)\right]\\
  &&
  +x\Gamma_Le^{i\chi}\Gamma_R(\Gamma_L+\lambda)\,.
\end{eqnarray}
For the present purpose, it is sufficient to determine the lowest
eigenvalue perturbatively in $x\ll 1$ and to assume
$\Gamma_L\ll\Gamma_R$. The counting statistics is obtained as
\begin{equation}
  \label{eq:cgf-twolevels}
  S(\chi)=-2x\Gamma_Lt_0\frac{e^{i\chi}-1}{2-e^{i\chi}}\,.
\end{equation}
This cumulant generating function reproduces, of course, correctly the
first cumulant $C_1=2x\Gamma_Lt_0$ and the noise $C_2=3C_1$,
corresponding to a Fano factor $F=C_2/|C_1|$ of $3$.\cite{cottet:prb}
The first interesting observation is that the current is \textit{twice}
the estimate of the simple thermal tunneling rate $x\Gamma_L$.
Furthermore, the noise corresponds to an \textit{effective charge}
$q_\mathrm{eff}=C_2/C_1=3$, which could be interpreted as a Poissonian
process of 3 charges. However, we obtain the for the higher cumulants
\begin{eqnarray}
  C_3=13 C_1 &,& C_4=75 C_1,\\
  C_5=541 C_1 &, & C_6=4683 C_1\,.
\end{eqnarray}
Notably, third (and higher) cumulants obey the relation
$C_n>q_{\mathrm{eff}}^{n-1}C_1$ and, consequently, the transport
statistics can not be explained by a simple Poissonian process in which
multiple charges of size $q_\mathrm{eff}$ are transfered. Note that this
would follow from a counting statistics 
$-(x\Gamma_L/q_\mathrm{eff})(\exp(iq_\mathrm{eff}\chi)-1)$. 

A simple physical picture emerges, if we expand the cumulant generating
function (\ref{eq:cgf-twolevels}) in terms of Poissonian processes. As a
result we find
\begin{equation}
  S(\chi)=-x\Gamma_L\sum_{n=1}^\infty \frac{1}{2^n}\left(e^{in\chi}-1\right)\,.
\end{equation}
The counting statistics is therefore a sum of independent Poisson
processes, in each of which a charge of $ne$ is transfered as signaled
by the factor $\exp(in\chi)-1$. Each process is weighted with a
probability $(1/2)^n$. This result for the statistics suggests the
following interpretation (see also Fig.~\ref{fig:cycles}). The lower
level is occupied most of the time, since it is well below the chemical
potentials of the two leads. Due to Coulomb blockade the other level
cannot be occupied at the same time (we assume here that the charging
energy is larger than the bias voltage). At a finite temperature, there
will be a small rate $\sim x\Gamma_L$ for the electron to hop out to the
left lead. Next, the dot will be filled again with an electron from the
right lead. However, as we are in the situation $eV_R>\epsilon_+$ both
levels will be occupied in the next step with the \textit{same}
probabilities of $1/2$.  In case the lower state is occupied we are back
to the initial state and one charge has been transfered in this cycle,
which is exactly what is happening in a single level quantum dot in the
thermally activated transport regime. However, if the electron has
tunneled onto the upper level, it can quickly tunnel out to the left
lead and the dot is empty again and another electron can tunnel from the
right terminal.  Each intermediate cycle occurs with a probability $1/2$
(since two states are available) and transfers an additional charge.
Therefore each full cycle has a probability $(1/2)^n$ and transfers $ne$
charge.

We now turn to the more general situation of a multi-level quantum
dot. The level structure we assume here is depicted in
Fig.~\ref{fig:cycles}b). As before the Coulomb energy is assumed to be
so large that only one of the levels can be occupied at the same
time. The levels are bunched into two groups, the lower group with $N_-$
levels and the upper group with $N_+$ levels. We will furthermore assume
that the tunneling rate into and from all levels inside a group are the
same. Physically this means that the maximal level spacing inside the
lower group is smaller than the temperature, so that the thermal
blocking factor $x=1-f(\epsilon_-)$ is the same for all levels. For the upper
group the requirement on the level spacing is less restrictive. We only
have to assume that the band width of the upper group is smaller than
the bias voltage. 

In principle we have to account now for $N_++N_1+1$ probabilities in the
Master equations, but due to the simplifications made above, we can
replace blocks containing 'upper'('lower') probabilities by a single
equation with the rate multiplied with a factor $N_{+(-)}$,
respectively. This procedure leads to the $\chi$-dependent rate matrix
\begin{equation}
  \hat M(\chi)=  \left(
  \begin{array}[c]{ccc}
    -x\Gamma_LN_- & 0 & \Gamma_RN_+\\
    0 & -\Gamma_LN_+ & \Gamma_RN_+\\
    x\Gamma_LN_-e^{i\chi} & \Gamma_LN_+e^{i\chi} & -2 \Gamma_RN_+
  \end{array}\right)\,,
\end{equation}
As before, it is sufficient to evaluate the CGF to lowest order in $x$. Let us
introduce $p=N_+/(N_-+N_+)$, which is the relative probability that an
electron tunneling from the right electrode occupies the upper
level. The counting statistics (in first order in $x$) is
\begin{equation}
  \label{eq:cgf2}
  S(\chi)=\lambda_0(\chi)t_0=-x\Gamma_Lt_0\frac{e^{i\chi}-1}{1-pe^{i\chi}}\,.
\end{equation}
This CGF gives the first cumulant $C_1=x\Gamma_Lt_0/(1-p)$ and the
higher cumulants are
\begin{eqnarray}\nonumber
  C_2 & = & \frac{1+p}{1-p} C_1,\\\nonumber
  C_3 & = & \frac{1+4p+p^2}{(1-p)^2}C_1,\\\nonumber
  C_4 & = & \frac{(1+p)(1+10p+p^2)}{(1-p)^3} C_1,\\\nonumber
  C_5 & = & \frac{1+26p+66p^2+26p^3+p^4}{(1-p)^4} C_1,\\\nonumber
  C_6 & = & \frac{(1+p)(1+56p+246p^2+56p^3+p^4)}{(1-p)^5} C_1\,.
\end{eqnarray}
We note that the average is larger by the factor $1/(1-p)$ than the
Poissonian estimate. Furthermore, higher cumulants obey the relation
$C_n>(C_2/C_1)^{n-1}C_1$,\cite{superproof} which shows the unusual characteristics of the
transport. Similar to the previous case, we can decompose the FCS into
multiple Poisson processes as
\begin{equation}
  \label{eq:cgf2a}
  S(\chi)=-x\Gamma_L (1-p)\sum_{n=1}^\infty p^{n-1}\left(e^{in\chi}-1\right)\,.
\end{equation}
The statistics is a sum of Poissonian processes of multiple charges,
weighted with probabilities $P_n=p^{n-1}(1-p)$. This suggests the
picture depicted in Fig.~\ref{fig:cycles}c). The possible Poissonian
processes result from a tunneling event with the small initial rate
$x\Gamma_L$.  This process is followed by $n-1$ fast cycles, in which an
electron tunnels through the dot with probability $p$. Finally the fast
cycle stops with a probability $1-p$. As can be seen from
Eq.~(\ref{eq:cgf2a}) processes of \textit{all} orders contribute to
the transport characteristics. A simplified description in terms of a
Poissonian process, in which an effective charge tunnels, can never
reproduce the observed cumulants. It is worth to emphasize, that only
the complete determination of all cumulants (or equivalently the full
counting statistics) can unambiguously identify this transport
mechanism. 

At this stage, we would like to remark that the situation, discussed
here, can only occur if the levels for zero bias are situated below the
Fermi level (assuming symmetric capacitive coupling). In case the levels
are initially above the Fermi level, the current and the noise show the
usual behaviors \cite{thielmann:03}. The corresponding counting
statistics can be obtained using the method of
Ref.~\onlinecite{bagrets:03} and turns out to be
\begin{equation}
  S(\chi)=\frac{\Gamma}{2}\sqrt{1+p_L(e^{i\chi}-1)+p_R(e^{-i\chi}-1)}
  -\frac{\Gamma}{2}\,. 
\end{equation}
Here $p_{L,R}$ are the respective probabilities of a tunneling event to
the left(right) and $\Gamma$ refers to the total tunneling rate. In the
regime of unidirectional transport $p_L\gg p_R$ and we obtain
\begin{equation}
  C_1=t_0\frac{\Gamma}{4}p_L\quad \mbox{and} \quad F=1-p_L/2\,.
\end{equation}
The Fano factor is always between $1/2$ for the symmetric case ($p_L=1$)
and 1 for the asymmetric case ($p_L\ll 1$). It follows also, that
higher-order cumulants are suppressed even further. This situation is
remarkably similar to transport through a noninteracting resonant level
in the limit of a bias voltage much smaller than the width of the
resonance.

The difference between super-Poissonian and sub-Poissonian statistics
discussed here has several interesting consequences.  First, we note
that the second cumulant in the super-Poissonian case is enhanced by a
factor $1+2N_+/N_-$. If for example $N_-=1$, the enhancement of the
noise is a direct measure of the number of excited states within the
transport window. Notably, the full information requires to study also
the third (and higher) cumulants, since higher cumulants are not related
to the second cumulant in a simple way.  These observations are
important for experimental studies of transport through molecules and
other quantum dots with an unknown internal level structures. As an
example we consider a molecule, in which it is not known, whether
transport takes place through the lowest unoccupied molecular orbital
(LUMO) or the highest unoccupied molecular orbital (HOMO), because the
position of the Fermi energy of the leads is not known. The current in
both cases shows a step at a threshold voltage.  However, the noise is
able to distinguish both situations. For transport through a multitude
of LUMO levels, the noise is always sub-Poissonian, see e.g.
\cite{setnoise,thielmann:03}.  However, transport through the HOMO falls
in the category discussed in this work. If there are excited levels in
the transport window, which cannot be occupied in equilibrium due to
Coulomb repulsion, we predict a super-Poissonian noise. Thus, noise
measurements may be used to distinguish transport through the HOMO or
the LUMO, which show the same average current-voltage characteristics.

In summary, we have pointed out the unusual noise characteristics of the
transport through a quantum dot with an internal level structure. A
super-Poissonian noise results from a combination of Poissonian
processes with multiple charges weighted with probabilities, depending
on the internal level structure. The obtained results allow to identify
the level structure, which is a useful information in molecular
electronic devices. We emphasize here, that only a study of the complete
statistics allows a determination of the transport mechanism underlying
the super-Poissonian shot noise.  While we have explored a rather
simplistic model of a quantum dot, further studies will reveal in more
detail what information can be obtained from current fluctuation
measurements in quantum dots and molecules with a complicated internal
level structure.

I acknowledge useful discussion with C. Bruder and A. Cottet. This work
was supported by the Swiss NSF, the NCCR \textit{Nanoscience} and the
RTN \textit{Spintronics}.


\begin{thebibliography}{99}


\bibitem{blanter:00} Ya.~M. Blanter and M. B\"{u}ttiker, Phys. Rep.
  \textbf{336}, 1 (2000).
  
\bibitem{nazarov:03} \textit{Quantum Noise in Mesoscopic Physics},
  edited by Yu. V. Nazarov (Kluwer, Dordrecht, 2003).

\bibitem{levitov:93} L. S. Levitov and G. B. Lesovik,  Pis'ma
  Zh. Eksp. Teor. Fiz. {\bf 58}, 225 (1993);
  %[JETP Lett. {\bf 58},  230 (1993)].
  L.~S. Levitov, H.~W. Lee, and G.~B. Lesovik,
 J. Math. Phys. {\bf 37}, 4845 (1996).

\bibitem{khlus:87}V.~A. Khlus, Sov. Phys. JETP \textbf{66}, 1243 (1987);%.
                                %\bibitem {lesovik:89}
  G.~B. Lesovik, JETP Lett. \textbf{49}, 592 (1989);%.
                                %\bibitem {buettiker:90}
  M. B\"{u}ttiker, Phys. Rev. Lett. \textbf{65}, 2901 (1990).

\bibitem{2e}
  M.~J.~M. de Jong and C.~W.~J. Beenakker, Phys. Rev. B {\bf 49},
  16070 (1994); T. Martin, Phys. Lett. A \textbf{220}, 137 (1996);
  M.~P. Anantram and S. Datta, Phys. Rev. B \textbf{53}, 16390 (1996);
  J. Torres and T. Martin, Eur. Phys. J. B \textbf{12}, 319 (1999);
  Fauchere ;
  J. Torres, T. Martin, and G.~B. Lesovik, Phys. Rev. B \textbf{63}, 134517 (2001);
  M. Schechter, Y. Imry, and Y. Levinson, Phys. Rev. B \textbf{64}, 224513 (2001);
  X. Jehl {\em et al.}, 
%  P. Payet-Burin, C. Baraduc, R. Calemczuk, and M. Sanquer,
  Phys. Rev. Lett. {\bf 83}, 1660 (1999);
  X. Jehl \textit{et al.}, Nature {\bf 405}, 50 (2000);
  A.~A. Kozhevnikov, R.~J. Schoelkopf, and D.~E. Prober,
  Phys. Rev. Lett {\bf 84}, 3398 (2000);
  K.~E.  Nagaev and M. B\"uttiker, Phys. Rev. B {\bf 63},
  081301(R) (2001).
  M.~P.~V. Stenberg and T~T. Heikkil\"a, Phys. Rev. B \textbf{66},
  144504 (2002);
  F. Pistolesi, G. Bignon, and F. W. J. Hekking, cond-mat/0303165
  (unpublished);
  F. Lefloch, C. Hoffmann, M. Sanquer, and D. Quirion,
  Phys. Rev. Lett. \textbf{90}, 067002 (2003).


\bibitem{2efcs} B. A. Muzykantskii and D. E. Khmelnitzkii,
  Phys. Rev. B {\bf 50}, 3982 (1994);
  W. Belzig and Yu.~V. Nazarov, Phys. Rev. Lett. {\bf 87}, 197006
  (2001); Phys. Rev. Lett. {\bf 87}, 067006 (2001);
  J. B\"orlin, W. Belzig, and C. Bruder, Phys. Rev. Lett. {\bf 88}, 
  197001 (2002);
  P. Samuelsson and M. B\"uttiker,
  Phys. Rev. Lett. \textbf{89},  046601 (2002);
  W. Belzig and P. Samuelsson,
  % \textit{Full Counting Statistics of Incoherent Andreev Transport}\\
  Europhys. Lett. \textbf{64}, 253 (2003).

\bibitem{cuevas:03}
  J.C. Cuevas and W. Belzig, Phys. Rev. Lett. {\bf 91}, 187001 (2003);
  G. Johansson, P. Samuelsson and A. Ingerman, Phys. Rev. Lett. {\bf 91}, 187002 (2003);
  J.C. Cuevas and W. Belzig, Phys. Rev. B  \textbf{70}, 214512 (2004); S. Pilgram
  and P. Samuelsson, cond-mat/0407029 (unpublished).
  
\bibitem{setnoise} A.~N. Korotkov, Phys. Rev. B \textbf{49}, 10381
  (1994); S.  Hershfield, J.~H. Davies, P. Hyldgaard, C.~J. Stanton, and
  J.~W. Wilkins, \textit{ibid.} \textbf{47}, 1967 (1993); U. Hanke,
  Yu.~M. Galperin, K.~A.  Chao, and N. Zou, \textit{ibid.} \textbf{48},
  17209 (1993); H. Birk, M.~J.~M. de Jong, and C. Sch\"{o}nenberger,
  Phys. Rev. Lett. \textbf{75}, 1610 (1995).%H. Birk, K. Oostveen, and C.
%Sch\"{o}nenberger, \ Rev. Sci. Instrum. \textbf{67}, 2977 (1996).

\bibitem{bagrets:03}
   D.~A. Bagrets and Yu.~V. Nazarov, Phys. Rev. B \textbf{67}, 085316
   (2003).


\bibitem{bulka}B.~R. Bulka, J. Martinek, G. Michalek, and\ J. Barnas,
Phys. Rev. B \textbf{60}, 12246 (1999); B.~R. Bulka, Phys. Rev. B \textbf{62},
1186 (2000).

\bibitem{cottet}    
  A. Cottet and W. Belzig,
%  \textit{Dynamical Spin-Blockade in a quantum dot with paramagnetic leads}\\
  Europhys. Lett. \textbf{66}, 405 (2004);
  A. Cottet, W. Belzig, and C. Bruder,
%  \textit{Positive Cross Correlations in a Three-Terminal Quantum
%     Dot with Ferromagnetic Contacts}\\
  Phys. Rev. Lett. \textbf{92}, 206801 (2004).

\bibitem{coupled}
G. Michalek and B. R. Bulka,  Eur. Phys. J. B {\bf 28}, 121 (2002);
G. Kie\ss lich, A. Wacker, and E. Sch\"{o}ll, Phys. Rev. B
\textbf{68}, 125320 (2003);
S.~S. Safonov% \textit{et al.}
, A.~K. Savchenko, D.~A. Bagrets, O.~N. Jouravlev, Y.~V. Nazarov, E.~H.
Linfield, and D.~A. Ritchie, Phys. Rev. Lett. \textbf{91}, 136801
(2003); O. Sauret and D. Feinberg, Phys. Rev. Lett. \textbf{92}, 106601
(2004).

\bibitem{blanter:99} 
  Ya.~M. Blanter and M.~B\"uttiker, Phys. Rev. B \textbf{59}, 10217 (1999).

\bibitem{sukhorukov} E.~V. Sukhorukov, G. Burkard, and D. Loss, Phys. Rev. B
  \textbf{63}, 125315 (2001).
  
\bibitem{buettiker:92} M. B\"{u}ttiker, Phys. Rev. B \textbf{46}, 12485
  (1992); for a review, see M. B\"{u}ttiker, in Ref.~\cite{nazarov:03}.

\bibitem{negdiff} Y.~P. Li, A. Zaslavski, D.~C. Tsui, M. Santos, and M.
  Shayegan, Phys. Rev. B \textbf{41}, 8388 (1990); E.~R. Brown, IEEE
  Trans. Electron Devices \textbf{39}, 2686 (1992); V.~V. Kuznetsov,
  E.~E. Mendez, J.~D. Bruno, and J.~T. Pham, Phys. Rev. B \textbf{58},
  10159 (1998); G. Iannaccone, G. Lombardi, M. Macucci, and B.
  Pelegrini, Phys. Rev. Lett. \textbf{80}, 1054 (1998);

\bibitem{cottet:prb}
  A. Cottet, W. Belzig, and C. Bruder,
  Phys. Rev. B \textbf{70}, 115315 (2004).

\bibitem{oppen} 
  J. Koch and F. von Oppen, cond-mat/0409667 (unpublished);  J. Koch, M.E. Raikh,
  and F. von Oppen, cond-mat/0501065 (unpublished).

\bibitem{jordan}
 A.~N. Jordan and E.~V. Sukhorukov,
 Phys. Rev. Lett. \textbf{93}, 260604 (2004).

\bibitem{thielmann:03} A. Thielmann, M.~H. Hettler, J. K\"{o}nig, and G.
  Sch\"{o}n, Phys. Rev. B \textbf{68}, 115105 (2003); cond-mat/0406647 (unpublished).

\bibitem{superproof} That this relation is obeyed for all cumulants can
  be seen by writing $C_2(\chi)=C_1(\chi)q(\chi)$, where
  $C_n(\chi)=-(-i\partial_\chi)^nS(\chi)$ and
  $q(\chi)=(1+pe^{i\chi})/(1-pe^{i\chi})$ yields the effective charge
  $q(0)=(1+p)/(1-p)$. Since all derivatives w.r.t. $i\chi$ of $q(\chi)$
  are positive, the chain rule yields $C_{n+1}(0)\geq q(0) C_{n}(0)$.

\end{thebibliography}
\end{document}